\documentclass[lettersize,journal]{IEEEtran}
\usepackage{amsmath,amsfonts}
\usepackage{algorithmic}
\usepackage{algorithm}
\usepackage{array}
\usepackage[caption=false,font=normalsize,labelfont=sf,textfont=sf]{subfig}
\usepackage{textcomp}
\usepackage{stfloats}
\usepackage{verbatim}
\usepackage{graphicx}
\usepackage{textcomp}
\usepackage{xcolor}
\usepackage{cite}
\usepackage{booktabs}
\usepackage[hyphens,spaces,obeyspaces]{url}
\usepackage[acronym]{glossaries}
\newacronym{aws}{AWS}{Amazon Web Services}
\newacronym{iot}{IoT}{Internet of Things}
\newacronym{lorawan}{LoRaWAN}{Long Range Wide Area Network}
\newacronym{lns}{LNS}{LoRaWAN Network Server}
\newacronym{cups}{CUPS}{Configuration and Update Server}
\newacronym{tls}{TLS}{Transport Layer Security}
\newacronym{vpc}{VPC}{Virtual Private Cloud}
\newacronym{api}{API}{Application Program
Interface}

\newcommand{\fc}[1]{\iftrue \textcolor{red} {\textbf{FC: #1}} \fi}

\usepackage{tikz}
\usetikzlibrary{shapes,arrows,positioning}

\begin{document}

\title{Monitoring the Venice Lagoon: an IoT Cloud-Based Sensor Network Approach}

\author{Filippo Campagnaro$^*$,~\IEEEmembership{Member,~IEEE,} Matin Ghalkhani$^*$, Riccardo Tumiati$^\ddag$, Federico Marin$^*$, Matteo Del Grande$^*$, Alessandro Pozzebon$^*$,~\IEEEmembership{Senior Member,~IEEE,} Davide De Battisti$^\S$, Roberto Francescon$^*$,~\IEEEmembership{Member,~IEEE,} Michele Zorzi$^{*\ddag}$,~\IEEEmembership{Fellow,~IEEE}

\IEEEauthorblockA{
		\small $*$ Department of Information Engineering, University of Padova, via Gradenigo 6/B, 35131 Padova, Italy }
  
\IEEEauthorblockA{
		\small $\ddag$ Wireless and More srl, Piazza Luigi da Porto 20, 35131 Padova, Italy }

\IEEEauthorblockA{
		\small $\S$ Department of Biology, University of Padova, Via U.Bassi 58/ B, 35131 Padova, Italy  }
  
\IEEEauthorblockA{
		\small $*$ \texttt{ \{campagn1,frances1,zorzi\}@dei.unipd.it
  alessandro.pozzebon@unipd.it, matin.ghalkhani@phd.unipd.it, federico.marin.2@studenti.unipd.it
  }} 
  
  \IEEEauthorblockA{
		\small $\S$ \texttt{ 
  davide.debattisti@unipd.it }} 

\IEEEauthorblockA{
		\small $\ddag$ \texttt{\{riccardo.tumiati,michele.zorzi\}@wirelessandmore.it}}  
}

\markboth{This manuscript is submitted for the IEEE JOE. The copyright for this content may be subject to change in the future without prior notice.}{}


\maketitle

\begin{abstract}
Monitoring the coastal area of the Venice Lagoon is of significant importance. While the impact of global warming is felt worldwide, coastal and littoral regions bear the brunt more prominently. These areas not only face the threat of rising sea levels but also contend with the escalating occurrence of seaquakes and floods. Additionally, the intricate ecosystems of rivers, seas, and lakes undergo profound transformations due to climate change and pollutants.

Employing devices like the SENSWICH floating wireless sensor presented in this article and similar measurement instruments proves invaluable to automate environmental monitoring, hence eliminating the need for manual sampling campaigns. The utilization of wireless measurement devices offers cost-effectiveness, real-time analysis, and a reduction in human resource requirements. Storing data in cloud services further enhances the ability to monitor parameter changes over extended time intervals.

In this article, we present an enhanced sensing device aimed at automating water quality assessment, while considering power consumption and reducing circuit complexity. Specifically, we will introduce the new schematic and circuit of SENSWICH which had changes in circuit and electronic aspects. Furthermore, we outline the methodology for aggregating data in a cloud service environment, such as Amazon Web Service (AWS), and using Grafana for visualization. 

\end{abstract}


\section{Introduction}

In the contemporary dynamic environment, a paramount challenge confronting us is climate change. The impact of planetary heating is experienced worldwide, with coastal and littoral zones exemplifying prominently its effects. These areas not only are at risk from escalating sea levels but also confront heightened occurrences of seaquakes and floods. Moreover, the complex ecosystems of rivers, seas, and lakes, encompassing crucial biodiversity hotspots such as the Natura 2000 protected areas~\cite{natura2000}, undergo profound alterations attributable to the repercussions of climate change and pollutants.

Among the promptly discernible outcomes of climate change in coastal regions is the elevation of sea levels. As the Earth's temperatures persistently rise, polar ice caps undergo melting, causing an expansion of the oceans. This expansion, coupled with the dissolution of glaciers and ice sheets, contributes to heightened sea levels. Coastal communities on a global scale are grappling with the ingress of seawater into their domains, resulting in coastal erosion, land depletion, and the displacement of populations.

The occurrence of  underwater earthquake and floods have been associated with climate change. The warming of ocean waters has the potential to destabilize tectonic plates, increasing the likelihood of seaquakes. Moreover, elevated temperatures can intensify rainfall events and, when combined with rising sea levels, may lead to devastating floods in coastal areas.

Natura 2000 constitutes a network of conservation zones within the European Union, comprising Special Areas of Conservation and Special Protection Areas designated under the Habitats Directive and the Birds Directive, respectively. This network encompasses both terrestrial and Marine Protected Areas~\cite{biodiversity}. Encompassing diverse ecosystems, including coastal habitats, these areas face a notable risk to biodiversity from the evolving climate. The elevation of sea levels has the potential to disturb the fragile equilibrium within these ecosystems, posing a threat to various plant and animal species.

The effects of climate change are exacerbated by pollution, as pollutants stemming from human activities, including industrial runoff and agricultural chemicals, enter rivers and oceans. These contaminants adversely affect aquatic life and add additional pressure on the fragile ecosystems of coastal and littoral regions.

Acknowledging the gravity of the circumstance, the European Union has initiated the European Biodiversity Strategy for 2030~\cite{eu-bio}. This ambitious endeavor strives to safeguard nature and counteract the deterioration of ecosystems. It endeavors to revive impaired ecosystems, prioritize biodiversity in all policy domains, and tackle primary factors contributing to biodiversity decline, such as climate change.

On a global scale, the United Nations has designated the Decade of Ocean Science for Sustainable Development (2021-2030)~\cite{un-od}. This program tries to enhance global collaboration in advancing scientific research and pioneering technologies that connect ocean science with societal requirements. It represents a crucial advancement in comprehending and tackling the challenges confronted by coastal and littoral regions.

Due to the complex nature and constant changes in coastal systems, efficiently monitoring them poses a significant challenge. Nevertheless, inventive solutions are arising. Intelligent sensors are being utilized to observe diverse environmental factors in aquatic ecosystems. These sensors provide real-time data, enabling the anticipation, administration, and alleviation of the impacts of climate change and pollution.

\begin{figure}[H]
    \centering
    \includegraphics[width=0.8\linewidth]{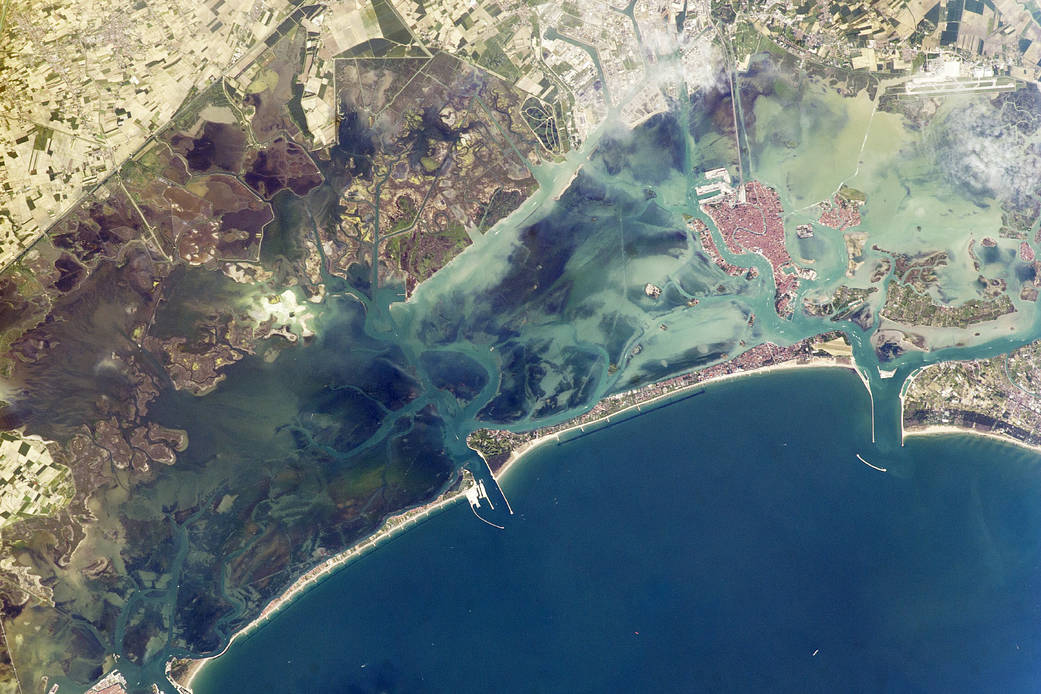}
    \caption{Lagoon of Venice, NASA, Aug 7, 2017.}
    \label{fig:venicelagoon}
\end{figure}

The Venice lagoon in Italy stands as a distinctive and intricate ecosystem that has been a focal point for researchers over an extended period. Illustrated in Figure~\ref{fig:venicelagoon}, this brackish and shallow water body features salt marshes and experiences significant changes in tides. Investigating the water parameters within this environment proves to be a challenging undertaking. 

The main contributions of this paper, which extends~\cite{metrosea2023}, are the presentation of the final architecture of a water monitoring platform for littoral areas, and the detailed description of the newly-designed advanced prototype of the sensory system and the cloud infrastructure for collecting the data, making it available to scientists and researchers.

This article explores advancements in networking architectures with a focus on the innovative Surface Node concept named SENSWICH. Beginning with a literature review in Section~\ref{sec:related}, we provide insights into existing research in the field. Section~\ref{sec:senswich} delves into the core of the contribution, presenting the design principles and theoretical foundations of SENSWICH. The implementation details, specifically the utilization of Amazon Web Services (AWS) Cloud Infrastructure, are discussed in Section~\ref{sec:aws}. Moving on to Section~\ref{sec:results}, we present results and engage in a comprehensive discussion, highlighting the performance metrics and implications of SENSWICH. The article concludes in Section~\ref{sec:conclusion} by summarizing key contributions and discussing potential future research directions.

\section{Related Works}\label{sec:related}

The examination of the marine ecosystem using smart sensor buoys has a long history that spans several decades. Initially developed for meteorological purposes, these buoys have transformed into essential instruments for investigating various marine factors. In recent years, the implementation of extensive ocean observatory systems in crucial places has brought about important changes in marine research.

Modern smart sensor buoys, as we recognize them today, have their origins in early meteorological research~\cite{garcia86}. Originally crafted to gather information on sea atmospheric conditions, these buoys offered valuable insights into weather patterns and climate.

The functionalities of smart sensor buoys also evolved with respect to the progression of technology. Researchers quickly grasped the potential of utilizing these buoys for investigating ocean currents~\cite{tabs}. This represented a noteworthy transition in their use, expanding beyond the realm of meteorology.

The role of ocean currents is pivotal in molding the Earth's climate, redistributing heat worldwide, and impacting weather systems. Gaining insight into these currents has become essential for various purposes, ranging from maritime navigation to research on climate change. Smart sensor buoys were modified to precisely measure the speed, direction, and depth of ocean currents. The information collected aided scientists in constructing detailed maps of currents, enhancing navigation safety for ships, and deepening our comprehension of how currents influence marine ecosystems.

Ocean observatories consist of positioned mooring systems equipped with sensor arrays capable of measuring various water parameters at different depths~\cite{themo}. These observatories allow scientists to gather data from oceanic regions that were previously inaccessible. Located in crucial areas, these observatories utilize smart sensor buoys fitted with sensors measuring parameters like Conductivity or Temperature, and Depth (CTD) probes, Acoustic Doppler current profilers (ADCPs), fluorometers, and other meteorological sensors. Together, these sensors collect data for the understanding of the oceanic dynamics and for the monitoring of the ecosystems health. CTD probes assess seawater properties such as salinity, temperature, and pressure, while ADCPs precisely measure ocean currents. Furthermore, fluorometers identify chlorophyll levels, aiding in the monitoring of phytoplankton populations. To ensure uninterrupted data collection, the buoys are equipped with solar panels and substantial batteries capable of storing over 30 MJ of energy, crucial for operating industrial-grade PCs onboard and supporting high-power communication systems~\cite{themo}. In general, maintaining the continuous operation of smart sensor buoys in remote ocean locations poses an operational challenge.


Smart sensor buoys serve not only as tools for data collection but also as essential contributors to environmental impact studies. Researchers leverage the data collected by these buoys to evaluate the health of marine ecosystems, monitor pollution levels, and study the impact of climate change. The increasing concern about the human activities on marine environments is addressed through the continuous data collection by smart sensor buoys, aiding in ongoing environmental impact assessments. Scientists utilize this data to track changes in water quality, temperature, and biodiversity, facilitating the identification and mitigation of the negative effects of pollution and climate change on marine ecosystems. Moreover, it empowers scientists to make knowledgeable choices regarding conservation initiatives and the sustainable management of resources.

The significant costs associated with installing and maintaining smart sensor observatory buoys, each with a price reaching several thousand US dollars, are justified by the advanced technology and extensive sensor arrays integrated into these nodes. Despite the substantial financial commitment required for their intricate design, advanced sensors, and deployment challenges in the ocean environment, this expenditure is justified by the abundant data they generate. This data provides valuable insights into the oceans, playing a crucial role in advancing marine science for the well-being of our planet.

In the field of environmental monitoring, the significance of coastal, river, and lagoon ecosystems cannot be overstated as they play a crucial role in preserving the delicate balance of global biodiversity. However, effectively overseeing these shallow water areas, where the depth rarely exceeds 5 meters, poses distinctive challenges. The complexity of these regions, marked by the convergence of multiple channels, various river mouths, and extensive salt marshes, makes deploying a single large observatory buoy less practical. Furthermore, the high costs associated with deploying and maintaining multiple buoy observatories render it an impractical option. In such situations, opting for a low-cost, easily maintainable monitoring system emerges as a more suitable alternative. Substantial efforts have been directed towards this goal in recent years, aiming to fill the gaps in our understanding of these vital ecosystems.

Furthermore, the work in~\cite{catania_sensore} presents another improvement in shallow-water monitoring with the development of an experimental test referred to as the Internet of Underwater Things (IoUT), as introduced in~\cite{petrioliUcomms}. The system proposed in this contribution consists of two essential components: an underwater node and a surface node, both equipped with sensing devices. What distinguishes this system is its communication method, which relies on acoustics. The data collected by the underwater unit is transmitted to the surface unit which, in turn, sends it to a shore server using the Long Range Wide Area Network (LoRaWAN) technology. This prototype, utilizing an industrial-grade acoustic modem designed for offshore applications, was primarily designed to illustrate the feasibility of the concept. It is important to note that such IoUT system is not intended for medium- or long-term installations but rather serves as a proof of concept for underwater monitoring in challenging shallow-water environments.

The SubCULTron initiative, outlined in~\cite{subcultron}, introduces an array of underwater and surface sensors and vehicles aimed at revolutionizing the understanding and exploration of the underwater environments. Crafted with precision and featuring cost-effective acoustic modems and WiFi modules, these sensor systems have the potential to reshape the landscape of coastal sensor networks.

A notable outcome of the SubCULTron project is the establishment of H2ORobotics \cite{h2orobotics}, an innovative startup affiliated with the University of Zagreb. H2ORobotics has successfully brought some of the sensor systems developed in the project to the commercial market. Particularly noteworthy is their introduction of a budget-friendly surface vehicle, marking a significant step forward in making advanced underwater and surface sensors more widely accessible globally.

\section{SENSWICH: The Surface Node }\label{sec:senswich}

In this section we describe the SENSWICH surface node, designed to measure diverse parameters including pH, electrical conductivity, turbidity, liquid level (with a conscientious consideration of ensuring the dependability of outcomes contingent upon the apt positioning of surface sensors in the aqueous milieu), dissolved oxygen, and temperature. 
Antecedent examinations of SENSWICH and its inaugural trials in~\cite{metrosea2023} unveiled its efficacy as a cost-effective, real-time monitoring device for climate change. Hence, the relevance of such inquiries extends beyond the efficacy of the apparatus, encompassing the substantive import of extensive data acquisition and protracted analytical scrutiny. Consequently, a decision was made to deposit and scrutinize the amassed data in the \gls{aws} cloud service.

Moreover, 
the improvements of the main board structure involved a shift from using two Arduino boards to a single Arduino. This change has been made possible by a new circuitry arrangement aimed at ensuring reliable results and energy efficiency. Ensuring electrical isulation among voltage-based sensors was also a key focus in the updated version of SENSWICH. These significant modifications collectively enhance the establishment of a stronger and more effective environmental monitoring system.

\subsection{System Requirement}
The peculiarity and heterogeneity of the Venice Lagoon, characterized by the presence of several micro-environments due to the convergence of multiple channels, the presence of various river mouths, and extensive salt marshes, makes the deployment a single large observatory buoy not effective to well characterize such a peculiar environment. In such scenarios, the deployment of a low-cost, easy-to-maintain monitoring system emerges as a more suitable alternative. The system we designed is composed of several low-cost low-power floating sensors, named SENSWICH, deployed in the whole lagoon, able to collect and transmit water measurements for several consecutive days with high granularity in time and space. In order to lower the power consumption, the system is kept off most of the time, and switched on every 15 minutes, only for the time needed to acquire the measurements from the sensors and send them to shore. Four samples per hour is indeed considered to be sufficient to characterize the environment. The GPS of SENSWICH, that is the most power demanding component of the node, will always be maintained off: it will be turned on only when specifically required by the operator, that from shore can send a command to get the node position and verify if it is still located in its deployment site.
Ideally, the distance between neighbor sensors will be around a few hundred meters, but a more dense deployment is envisioned in hot-spot areas.
The high activity of sea life in the lagoon leads to water sensors becoming eventually encrusted with bio-fouling. This calls for periodic maintenance to clean the sensors: during this operation the battery will also be replaced. According to our previous knowledge, maintenance is required every sixty days, hence, we can dimension the battery capacity to provide an autonomy of about two months. We went through this solution instead of solar panels mainly for two reasons: 
\begin{itemize}
    \item extending the battery life for more than the time the sensors can be fully operational will only make the system more complex and expensive without providing any evident benefit;
    \item given the high number of fishermen, tourists and people moving across the lagoon, the use of floating systems should be as minimal as possible to discourage them them approaching it: the presence of a solar panel would easily attract the attention of people that may dangerously get close to it.
\end{itemize}

\begin{figure}[H]
\centering
\includegraphics[width=0.6\linewidth]{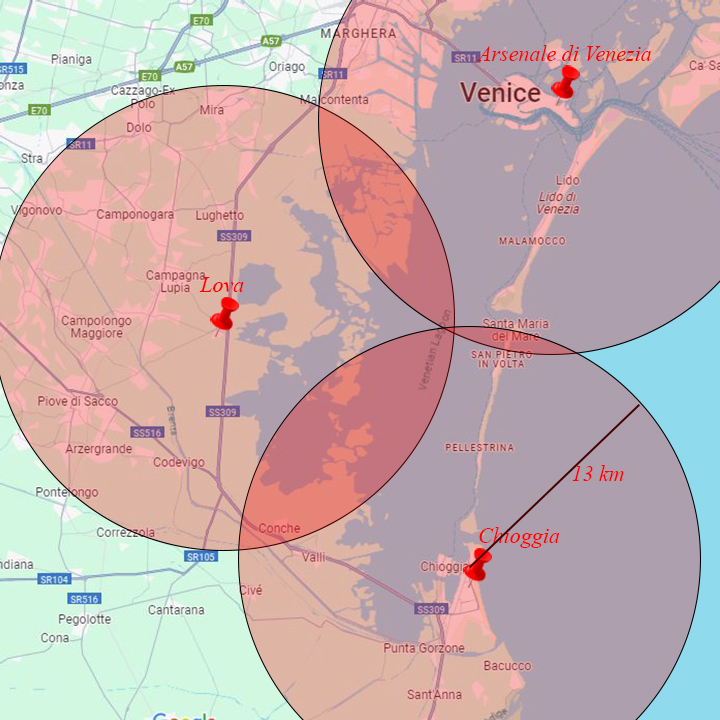}
\caption{Expected coverage area of three  \gls{lorawan} gateways.}
\label{fig:maps}
\end{figure}

To ensure extensive connectivity with a low-power transmitter, we opted to integrate a \gls{lorawan} transceiver into the sensor. This transceiver enables transmission to a shore gateway situated up to 10-15 km away from the sensor. This approach allows to effectively cover nearly the entire Venice lagoon with just three gateways. (Please, refer to Figure~\ref{fig:maps} for a visual representation of the coverage area.)

\subsection{Sensor Selection}

The compendium of sensors comprises:
\begin{itemize}
\item Analog Industrial pH Sensor/Meter Pro Kit V2 (SEN0169-V2);
\item Analog Electrical Conductivity Sensor/Meter K = 10 (DFR0300-H);
\item Analog Turbidity Sensor (SEN0189);
\item Waterproof DS18B20 Temperature Sensor Kit (KIT0021);
\item Analog Dissolved Oxygen Sensor / Meter Kit (SEN0237-A);
\item Photoelectric High Accuracy Liquid Level Sensor (SEN0205).
\end{itemize}

Together with these sensors, the system integrates also a TEL0094 GPS Module with Enclosure. Additionally, there is a shift from the previous model, now utilizing only one Arduino unit in the current setup. This decision is made to balance power conservation and circuit simplification. The chosen Arduino for essential tasks like data acquisition and transmission is the Arduino MKR WAN 1310 board. This device combines a SAMD21 Cortex-M0+ 32-bit low-power ARM MCU with a Murata CMWX1ZZABZ LoRa chip, working across the 433, 868, and 915 MHz frequency bands. The MCU, equipped with 7 Analog input pins and 8 Digital I/O pins, facilitates the efficient connection of all the sensors to a single board.

\begin{figure*}[!t]
    \centering
    \includegraphics[width=1\textwidth, trim=2cm 1.5cm 2cm 1.5cm, clip]{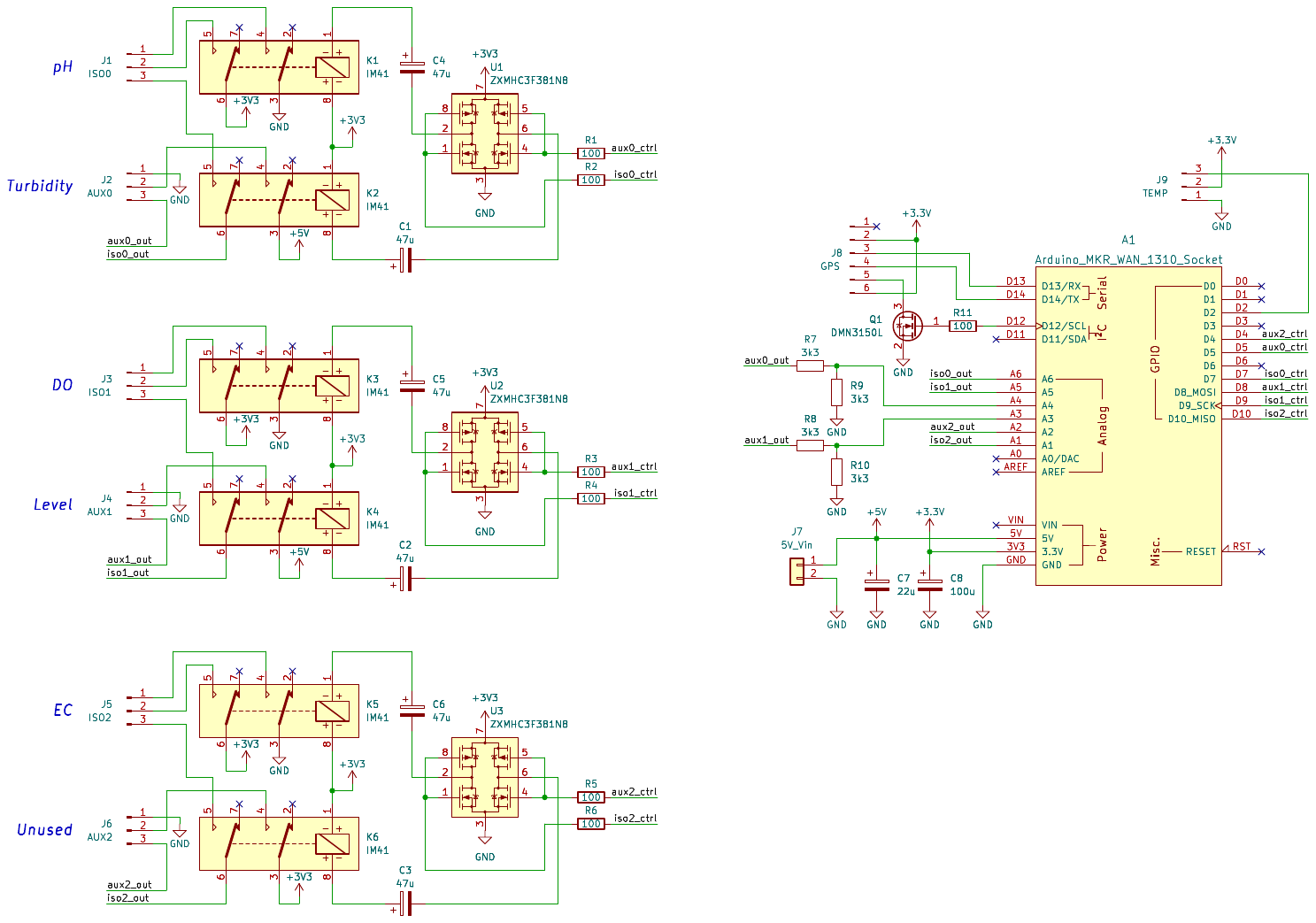}
    \caption{Schematic of SENSWICH with one Arduino (all relays initially are in reset mode).}
    \label{fig:schematic}
\end{figure*}

\begin{figure}[h]
\centering
\includegraphics[width=0.8\linewidth]{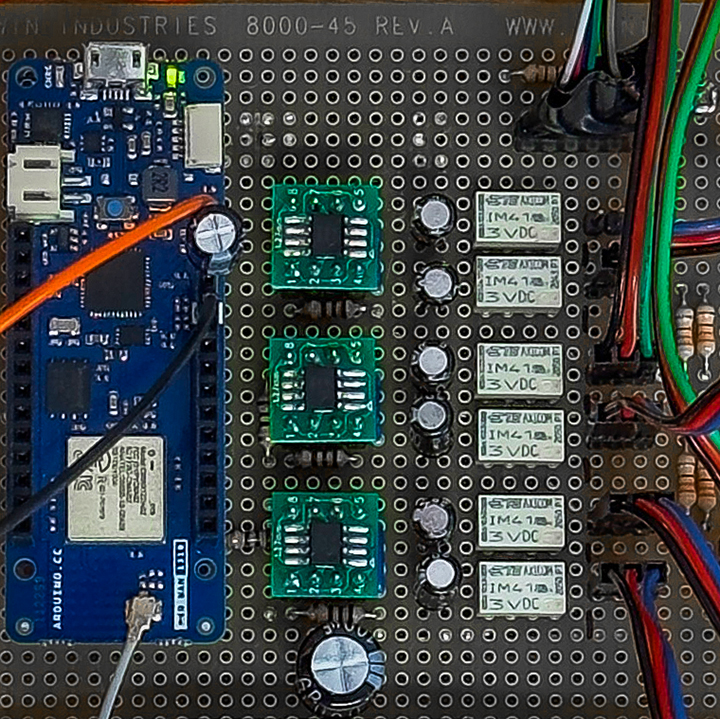}
\caption{Circuit of SENSWICH with one Arduino.}
\label{fig:circuit}
\end{figure}

\subsection{Data Measurement and Transmission Utilizing SENSWICH}\label{sec:senswich_tx}


In previous studies, we utilized two Arduinos instead of one, as detailed in \cite{metrosea2023}. To have an improvement in power consumption and also less complexity, we decided to reconfigure SENSWICH. After improving programmed codes and conducting electronic investigations, we wanted to be sure that there could be no interference between sensors inside a salty environment which may work as a conductor. To avoid having this type of interference, we started to investigate over different types of electrical isolators.

To create electrical isolation in our circuit, we studied different solutions. Initially, we analyzed existing isolation and assessed their performance based on some features like the manufacturing companies, signal type, zero-load input current, sleep current, and price (Table \ref{tab:ots_isolators}). 
\begin{table}[H]
    \centering
    \caption{Commercial solutions for electrical isolation.}
    \label{tab:ots_isolators}
    \begin{tabular}{cccccc}
        \hline
        \shortstack{Manufacturer} & Signal type & \shortstack{Zero load\\input current} & \shortstack{Sleep\\current} & Price \\
        \hline
        Atlas Scientific & \shortstack{UART, I\textsuperscript{2}C,\\SMBus} & 20 mA & 3.8 mA & \$24.99 \\
        Atlas Scientific & UART, I\textsuperscript{2}C & 22 mA & 2.6 mA & \$28.99 \\
        Atlas Scientific & Analog & 15.7 mA & --- & \$25.99 \\
        DFRobot & Analog & 75 mA & --- & \$19.90 \\
        DFRobot & I\textsuperscript{2}C & 15 mA & --- & \$16.00 \\
        Our Solution & Analog &  $\approx 0$ &  $\approx 0$ & $\approx \$29.40$ \\
        \hline
    \end{tabular}
\end{table}

As a result, we opted against using these power isolators due to three drawbacks listed as follows.

\begin{enumerate}
    \item As indicated in Table \ref{tab:ots_isolators}, the existing brands compatible with the SENSWICH circuit exhibited excessive power consumption, which is a critical consideration in our project~\cite{metrosea2023}. Given that even tens of milliamperes are significant in this context, these isolators were deemed unsuitable.
    
    \item Their efficiency did not meet our requirements, considering the high power consumption and the fact that they only separated electrical signals, a task achievable with alternative methods discussed later in this paper.
    
    \item Some isolators consumed power even during sleep mode, which is unfavorable for SENSWICH, as biologists require measurements of various water parameters every few minutes (e.g., every ten or twenty minutes).
\end{enumerate}

Hence, we decided to proceed with low-power relays: their arrangement in the SENSWITCH circuit is presented in Figure~\ref{fig:schematic}, 
 visually represents in Figure~\ref{fig:circuit}.
The circuit employs DPDT single-coil-latching micro-relays, specifically the IM41TS model manufactured by ``TE Connectivity." It is a power efficient relay and, according to its datasheet~\cite{datasheet}, the coil power consumption during switching is around 140 mW in standard models, 100 mW in high sensitive models, and 50 mW for ultra high sensitive models. As it performs switching in 1-3 ms, we can consider its contribution to the average power consumption as negligible. These components are designed for precise analog signal switching and offer several advantages:

\begin{itemize}
    \item compact size (10x6 mm), occupying only 60 mm\textsuperscript{2} of board space;
    \item excellent contact stability and long life due to gold-plated PdRu contacts, resulting in a connection resistance of less than 50 m$\Omega$ and a minimum electrical endurance of approximately 10\textsuperscript{6} switching operations—sufficient for the expected operational life of the node;
    \item low coil excitation voltage of 3 V, eliminating the need for complex voltage translation circuitry;
    \item hermetically sealed case to prevent performance degradation from moisture and contaminants in the atmosphere;
    \item affordable, with a cost of around \$3.5 per piece for single quantities.
\end{itemize}

To generate bidirectional current pulses for the relay coil configuration, we used a discrete MOSFET half-bridge driver with a series capacitor. The selected MOSFETs (U1 through U3 in Figure~\ref{fig:schematic}) have a low gate threshold voltage, enabling direct logic-level drive from an Arduino digital output pin. This eliminates the need for an external gate driver, reducing cost, complexity, and power consumption.

For complete electrical isolation of electrochemical sensors (pH, EC, and DO), all three connections from the signal conditioning module (VCC, GND, and analog output) must be interrupted. Thus, two separate relays with three mechanical contacts are utilized. The unused contact in this configuration switches the power terminal to one of the remaining sensors (turbidity and liquid level), which do not require complete electrical isolation since there is no electrical contact with water.

In general, this solution meets the isolation requirements for sensors while preserving signal integrity and consuming virtually zero idle power after the switching transient (which requires a pulse of 10~mA that lasts 1~ms). Additionally, it requires fewer than four external components per acquired parameter, reducing manufacturing cost and the risk of malfunctions during operation.

We equipped SENSWICH with 8 Li-Ion batteries according to the specifications in Table~\ref{tab:senswich_info}. Subsequently, we employed an LM2596 regulator to convert the voltage from nominal voltage 7.4 V to approximately 5.25 V, ensuring a stable power supply for the Arduino. Powering the Arduino at 5.25 V is crucial for optimal sensor functionality, and specifically, the turbidity and liquid level sensors require a 5 V input voltage. 

\begin{table}[h]
    \centering
    \caption{Information of Power Source of Senswich}
    \label{tab:senswich_info}
    \begin{tabular}{lr}
        \toprule
        \textbf{Parameter}              & \textbf{Value} \\
        \midrule
        Cell capacity (mAh)            & 3200           \\
        Cell energy (Wh)               & 11.5           \\
        Nominal cell voltage (V)       & 3.7            \\
        Number of series cells         & 2              \\
        Number of parallel cells       & 4              \\
        \bottomrule
    \end{tabular}
\end{table}

The Arduino is programmed to read the sensor data and send the sampled values via LoRa at regular intervals of 10 minutes, while remaining in the lowest power sleep mode available (called ``Standby mode" of the SAMD21 microcontroller) during periods of inactivity. An acquisition cycle is initiated by an interrupt from the Real-Time Clock (RTC) module, which ensures a constant sampling period, after which all the sensor data is read out following a procedure outlined later. Subsequently, the raw values are corrected using calibration constants stored inside the flash memory of microcontroller, organized into a single packet following the CayenneLPP structure. At the end, this single packet is sent to AWS over LoRa, afterwards the Arduino goes back to standby mode.

Because of the relay contact arrangement shown in Figure~\ref{fig:schematic}, a specific sequence of operations must be carried out. As an example, in order to read pH and turbidity, firstly relay K1 is switched to the ``set" position, powering the pH probe and the associated signal conditioning module. During the subsequent 80-second period required for this sensor to stabilize, the turbidity sensor is kept off to preserve energy. Next, relay K2 is set, simultaneously connecting the pH output to the analog input of the Arduino and powering the turbidity sensor, which will present stable readings after about 10 seconds. At the end, the voltage output from each sensor is read by the ADC, averaging over 256 samples to further reduce noise, and both relays are reset. This procedure is repeated for DO, liquid level, EC and the unused input J6.

Regarding the GPS feature, the activation of sensor activity begins by transmitting a ``gps" string in base64 encoding within the AWS Payload segment. It is crucial to note that the GPS sensor stays active for a maximum duration of 5 minutes (if it can find the location in less than that time, it is going to be turned off). After that, it enters a sleep state. If the sensor encounters challenges in determining its location due to unfavorable weather conditions or a displaced device, a new activation request needs to be sent through a designated message.

\section{Cloud Infrastructure}\label{sec:aws}
\begin{figure*}[!t]
\centering
\includegraphics[width=\textwidth]{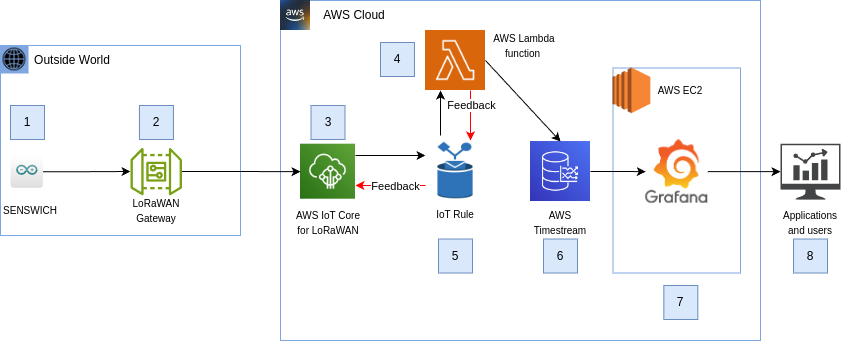}
\caption{Cloud architecture used to collect, store, and visualize the data.}
\label{fig:aws_architecture}
\end{figure*}

This section presents the Cloud architecture based on \gls{aws} used to store and visualize the data collected by SENSWICH. The design of the infrastructure will be discussed, describing the core services involved, from the data acquisition to the final visualization tools.

\subsection{Design of the Cloud Architecture}

Modern cloud computing platforms offer a wide range of services such as computing power, storage, and databases that can be used to build and deploy various types of applications, including IoT sensor networks.
One of the key advantages of using a cloud-based solution for an IoT sensor network is its scalability. Such platform usually allows for the easy addition and removal of resources as needed, which is important for a sensor network that may need to handle varying levels of data traffic. Moreover, a cloud platform also provides a range of security features, such as identity and access management, encryption, and network isolation, which are essential for protecting sensitive IoT sensor data. Additionally, it provides a wide range of services for data storage and management

For instance, AWS~\cite{aws-cloud} offers a variety of services specifically tailored for IoT, such as AWS IoT Core for device connectivity, AWS Greengrass for local device computing, and AWS IoT Analytics for data processing and analysis. Additionally, in terms of data storage and management, AWS includes Amazon Timestream, Amazon S3, Amazon DynamoDB and Amazon Elasticsearch, which can be used to store and analyze large amounts of sensor data.

The proposed solution for the Cloud architecture and \gls{iot} sensor network, as illustrated in Figure~\ref{fig:aws_architecture}, is composed of two main parts: the Outside world and the AWS cloud~\cite{aws-cloud}. The Outside world includes all the IoT Arduino endpoints and the \gls{lorawan} gateway, which establishes a secure connection to the \gls{aws} IoT Core for \gls{lorawan}~\cite{iot-core-lorawan} using a double channel and certificate-based authentication. The Arduino then encodes the water data and creates a payload using the CayenneLPP format~\cite{cayenne-lpp}. This payload is added to the LoRa packet and sent to the gateway.

The AWS cloud is composed of the AWS IoT Core for \gls{lorawan}, which receives the \gls{lorawan} packets from the gateway and triggers an AWS rule to handle these packets thanks to a topic in MQTT protocol~\cite{mqtt-protocol}. A SQL query inside the AWS rule is then used to initiate an AWS Lambda function, which decodes the data received from Base64 to CayenneLPP and then from CayenneLPP to ASCII cleartext. The packets are then modified and augmented with the new decoded data, and finally stored into the Amazon TimestreamDB using Python SDK, making it available for visualization through the Grafana service running on an AWS EC2 virtual machine. AWS Lambda function gives a feedback to the AWS rule on the operation success: if the decoding and storing processes are completed correctly, the packet is republished in \textit{lorawan} MQTT topic, otherwise in \textit{lorawan/error} MQTT topic, allowing the operator to understand what is happening in real-time.

Inside the Grafana service, users can be added with different permissions and groups and each user can use the admin preconfigured dashboard with tables and graphs to visualize quasi real-time data or create their own dashboards and query the Amazon Timestream database to retrieve the data. This allows for flexibility and ease of use in monitoring and analyzing the sensor data.

\begin{figure}[!t]
\centering
\includegraphics[width=3.5in]{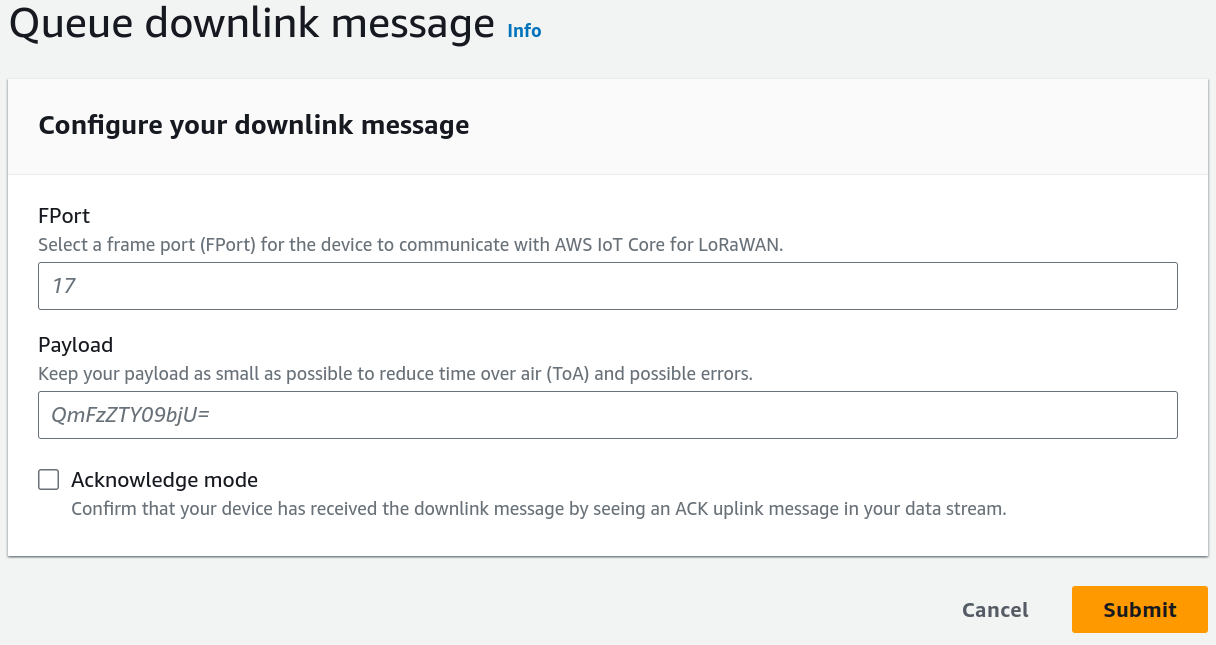}
\caption{Queue downlink message window.}
\label{fig:queue_downlink}
\end{figure}

This is the general operating mechanism for the common water measures, like temperature, PH, turbidity and the others. The approach is different to retrieve the GPS position of SENSWICH, which is not sent periodically and requires a manual action performed by the operator. In this case, the administrator wanting to get this information has to enqueue a downlink message to the sensor in order to activate GPS and receive the position in next periodic data transmission. This operation can be easily done with the AWS console (Figure~\ref{fig:queue_downlink}), where the operator can insert the FPort field of the application and the base64 encoded payload of the packet.

\subsection{LoRaWAN network server}
AWS IoT Core~\cite{iot-core} is a fully managed platform that enables the connection, management, and processing of data from IoT devices. It provides a set of services to connect and manage devices, including device registration, device provisioning and device management, allowing businesses to focus on creating innovative solutions and applications that leverage the data generated by IoT devices. Additionally, it includes a message broker that allows devices to securely and efficiently send and receive messages.

AWS IoT Core allows for the processing and analysis of data from devices using services such as AWS IoT Analytics and AWS IoT Events. These services enable the creation of complex workflows for data processing, analysis, and triggering of actions. It also enables integration with other AWS services, such as AWS Lambda, Amazon S3, Amazon DynamoDB, and Amazon Kinesis, providing a wide range of options for storing, analyzing, and acting on device generated data. AWS IoT Core also provides a secure communication channel between the base station and the cloud using the \gls{tls} protocol, which ensures that the data is transmitted securely.

In addition, AWS IoT Core also provides security features such as authentication, authorization, and access control to ensure the secure communication between devices and the cloud. It also supports the integration with other security mechanisms such as \gls{vpc} and PrivateLink.

AWS IoT Core for \gls{lorawan}~\cite{iot-core-lorawan} is a service that allows to connect and manage \gls{lorawan} gateways and end-point devices on AWS. The service enables to securely connect, provision and manage devices using the \gls{lorawan} protocol. The service also allows to process and route device data to AWS services and to create IoT applications for \gls{lorawan} devices. \gls{cups} and \gls{lns} are advanced features that can be used with the MultiTech Conduit® IP67 Base Station and AWS IoT Core to enhance the functionality and security of \gls{lorawan} networks.

\gls{cups}~\cite{cups} is a feature that allows the base station to support additional protocols beyond the standard LoRaWAN protocol. These protocols can be used to implement custom application layer functionalities, such as over-the-air firmware updates, device management and secure communication between devices and the network server.

\gls{lns}~\cite{lns} is a service that allows the base station to communicate with a network server using a secure and efficient protocol. This enables features such as device management, security and data management. \gls{lns} also provides a secure communication channel between the Base Station and the cloud using the Transport Layer Security protocol, which ensures that the data is transmitted securely.

The connection between the \gls{lorawan} gateway and AWS IoT Core for \gls{lorawan} is established using the standard \gls{lorawan} protocol and the \gls{lorawan} \gls{api}. The gateway connects to the AWS IoT Core for \gls{lorawan} using a unique device identity and security credentials, such as a certificate or key. This enables the gateway to securely send and receive data to and from the AWS IoT Core for \gls{lorawan} service. Once the connection is established, the gateway can send \gls{lorawan} packets containing sensor data from the Arduino endpoints to the AWS IoT Core for \gls{lorawan} service. The service then triggers rules that are used to process the data and route it to other AWS services such as AWS Lambda or Amazon TimestreamDB. This allows to create IoT applications that can process and analyze sensor data in real time.

AWS IoT Core for \gls{lorawan} uses the standard \gls{lorawan} architecture~\cite{lorawan-architecture} which includes three main components: the end devices, the gateway, and the \gls{lns}. The connection gateway, also known as the \gls{cups}, acts as the bridge between the end devices and the network server. End devices, such as the Arduino endpoints in the proposed solution, have sensors that collect data and send them to the gateway. These devices communicate with the gateway using the \gls{lorawan} protocol and the unique device identity and security credentials. 

The gateway, also known as the Basic Station Gateway~\cite{basics-station} receives the data from the end devices and forwards it to the network server; it also receives messages from the network server and forwards them to the end devices, allowing the transmission of actuation commands, like the GPS request in our case.

The Network Server is the central component of the \gls{lorawan} network, and is responsible for managing the communication between the gateways and the end-devices, for managing the security, the access control and the device management. \gls{lns} connects to the gateway using the standard \gls{lorawan} protocol and the unique gateway identity and security credentials. In the case of AWS IoT Core for \gls{lorawan}, the \gls{lns} is a fully managed service that allows to securely connect, provision, and manage devices. It also allows to process and route device data to AWS services, and to create IoT applications for \gls{lorawan} devices. In summary, the AWS IoT Core for \gls{lorawan} connection gateway consists of the \gls{cups} and \gls{lns}: in this architecture the end devices send the data to the Basic Station Gateway which then forwards it to the Network Server managed by AWS IoT Core for \gls{lorawan}. This allows for secure and efficient communication between the end devices and the network server and enables processing and routing of device data to AWS services for further analysis.

\subsection{Data processing and memorization}
AWS IoT Rules~\cite{iot-rules} is a service that enables the creation of rules that automatically perform actions on data from IoT devices when specific conditions are met. These rules are used to route, filter, and process data from IoT devices at the edge, in the AWS IoT Core, or in other AWS services.

AWS IoT Rules allows for the easy creation of rules that perform actions such as sending data to Amazon S3, Amazon DynamoDB, or other services for storage and analysis, triggering an AWS Lambda function, or sending a message to an Amazon SNS topic. Additionally, rules can be used to forward data to other AWS services like AWS Greengrass, AWS Kinesis, AWS Elasticsearch, and more.

Custom actions can also be implemented using AWS Lambda functions in multiple languages. Messages sent to the AWS IoT message broker trigger AWS IoT Rules. The messages are filtered and processed by the rule engine. The rules can be applied to all messages or specific messages based on topic, source, or payload.

AWS IoT Rules is a powerful service that allows for easy processing and analysis of data from IoT devices and the triggering of actions based on specific conditions. This service can be used to build a wide range of IoT applications and use cases, including smart homes, industrial automation, and predictive maintenance.

AWS Lambda~\cite{aws-lambda} is a serverless compute service that runs code in response to events and automatically manages the underlying compute resources. With Lambda, users can run code for virtually any application or backend service, all with zero administration. Users simply upload the code and Lambda takes care of everything required to run and scale their code with high availability. They can set up the code to automatically trigger it from other AWS services or call it directly from any web or mobile app.

AWS Lambda supports multiple programming languages including Java, Python, C\#, Go, and Node.js. It can be used to build event-driven applications and microservices, process and analyze data, and run backend services for mobile and web applications. It operates on a pay-per-use basis, with no charge when the code is not running. It also allows for automatic scaling based on incoming requests or other metrics, eliminating the need for provisioning and scaling servers, and integrates with other AWS services.

Integration with other AWS services, such as Amazon S3, Amazon DynamoDB, and Amazon SNS, allows for the building of powerful end-to-end applications. This makes AWS Lambda an attractive option for building event-driven and serverless applications due to its scalability and cost-effectiveness.

Amazon Timestream~\cite{aws-timestream} is a fully managed time series database service that enables users to store, process, and analyze time-stamped data at scale. It is designed to handle high write and query loads, and optimized for storing and analyzing time-stamped data, such as IoT sensor data, application logs, and clickstream data.

Users can create a database and define tables within it, where each table can have multiple dimensions and measures, allowing for easy data modeling and analysis. Timestream also provides a SQL-like query language to easily query and analyze data.

Timestream integrates with other AWS services, such as Amazon CloudWatch, AWS IoT
and AWS Lambda, making it easy to collect, store, and analyze time-stamped data from various sources. A retention period can be set to automatically expire data that is older than a certain number of days, reducing storage costs and allowing for better management of data over time.

\subsection{Data access and visualization}
AWS Elastic Compute Cloud (EC2)~\cite{aws-ec2} provides a platform to create virtual servers that can be used to run applications as if they were deployed in a local datacenter. The power of this AWS service is its flexibility: it is possible to provision the virtual machines with the hardware needed, equipping it with the required CPU, memory, storage and network interfaces.

When an operator creates an EC2 instance, it first has to choose the type of machine specifically from the use it has to accommodate, for instance there are general purpose machines, storage optimized machines, compute optimized ones and so on. The instance can be prepared with an Operating System (all the most common ones) and applications installing a predefined image or a custom one. Then, the external environment of the machine has to be configured, so the region where it is physically located and the \gls{vpc} on which it resides, configuring also its network interfaces accordingly. Moreover, the best security options for the specific use-case can be chosen and, finally, there is the possibility to scale the service in order to avoid general failures and manage the different loads.

In our infrastructure (Figure~\ref{fig:aws_architecture}) we just deployed one virtual server with Grafana application inside, which is publicly available in the Internet. Grafana~\cite{grafana} is a platform used for live monitoring and observabililty of data incoming from different sources, depending on the background solution implemented. As we know, in our specific case data are fetched from an AWS Timestream database hosted in the cloud. Grafana simplifies the setup and management of customizable dashboards and panels used to visualize the information for multiple teams or projects. Alerting and notifications feature allows users to set up alerts and notifications that trigger when specific conditions are met, such as when a metric exceeds a certain threshold. Role-based access control provides granular access controls that allow to specify which users have access to dashboards and panels. Data security is provided by a number of security features, such as encryption, to help protect user data.

\section{Results and Discussion}\label{sec:results}

\begin{figure}[h]
\centering
\includegraphics[width=1\linewidth]{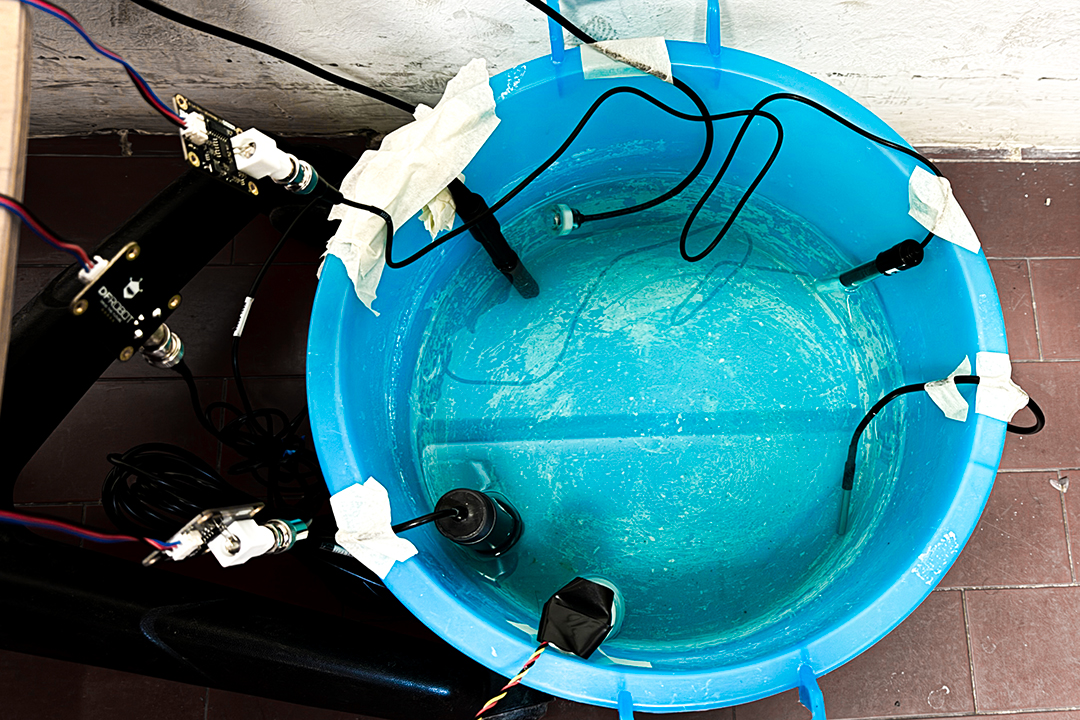}
\caption{Setup of sensors in a bucket.}
\label{fig:container}
\end{figure}

\begin{figure}[h]
\centering
\includegraphics[width=1\linewidth]{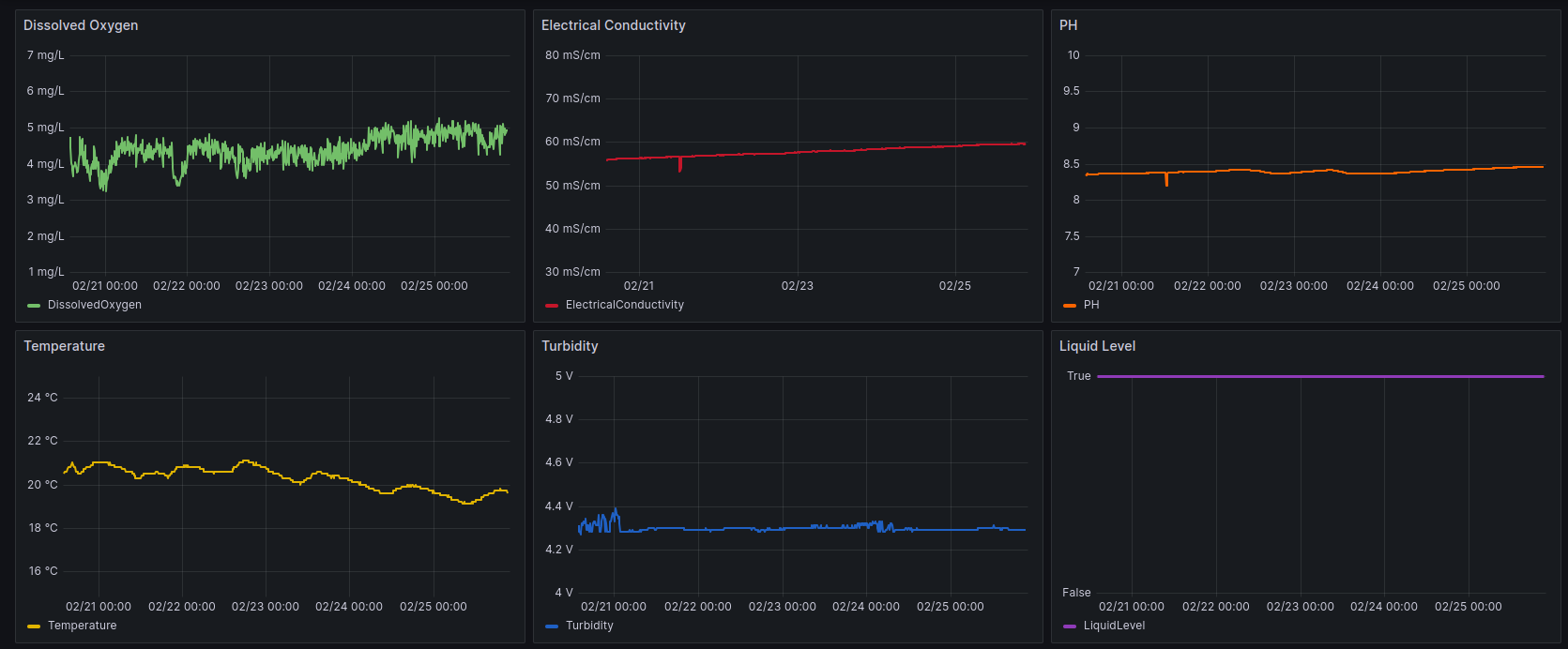}
\caption{Visualization in Grafana.}
\label{fig:grafana}
\vspace{0.5cm}
\centering
\includegraphics[width=1\linewidth]{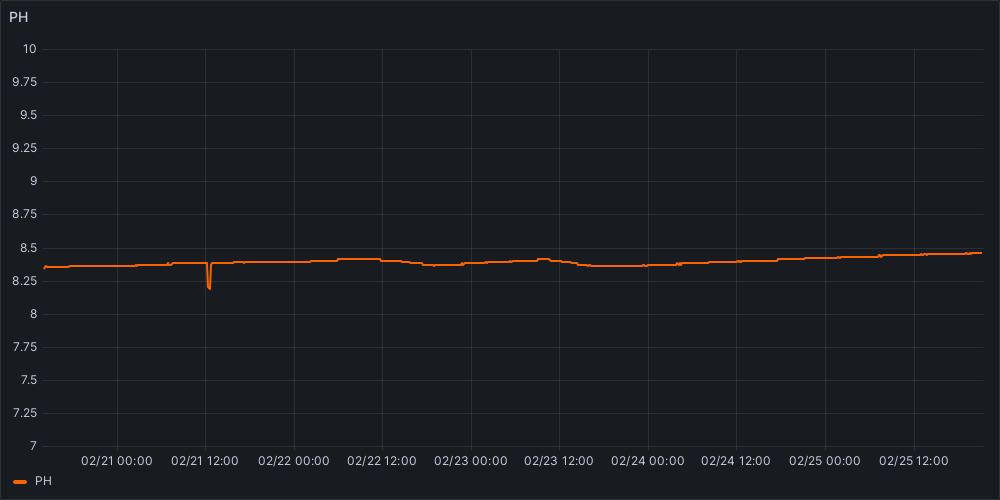}
\caption{pH sensor results.}
\label{fig:pH}
\vspace{0.5cm}
\centering
\includegraphics[width=1\linewidth]{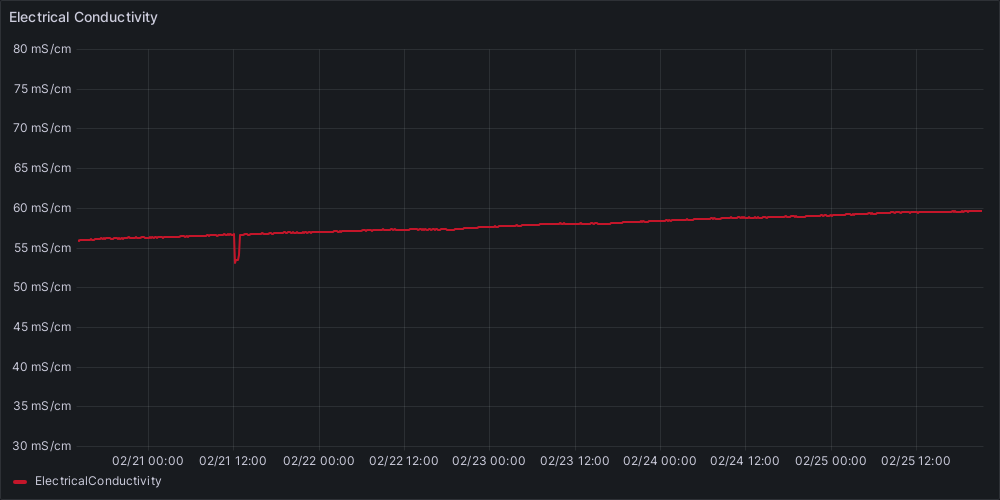}
\caption{Electrical conductivity sensor results.}
\label{fig:conductivity}
\vspace{0.5cm}
\centering
\includegraphics[width=1\linewidth]{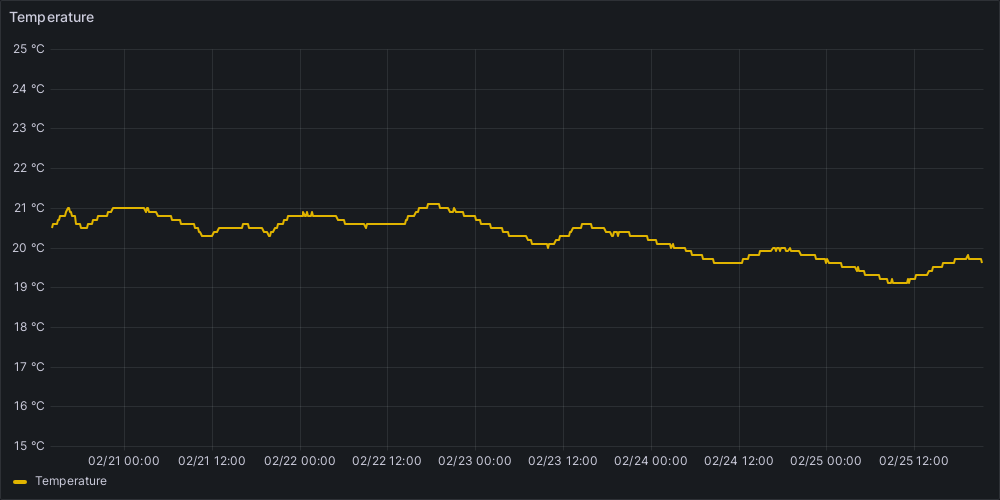}
\caption{Temperature sensor results.}
\label{fig:temperature}
\end{figure}

For testing purposes, we placed all sensors in a bucket filled with saline solution, as shown in Figure \ref{fig:container}. We utilized Grafana to integrate and display data from various sensors simultaneously, as shown in Figure \ref{fig:grafana}. After setting up the gateway and incorporating the Arduino into our AWS infrastructure, we successfully transmitted the data to Grafana. Subsequently, we initiated the data acquisition process and we observed reliable results from all sensors over an extended period, roughly spanning from February 20, 2024, to February 26, 2024.

\begin{figure}[h]
\centering
\includegraphics[width=1\linewidth]{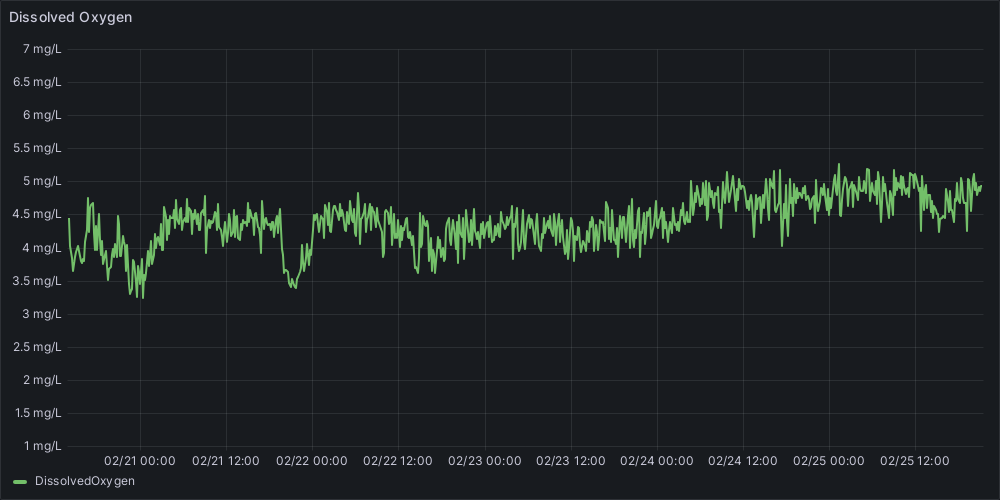}
\caption{Dissolved oxygen sensor results.}
\label{fig:oxygen}
\vspace{0.2cm}
\centering
\includegraphics[width=1\linewidth]{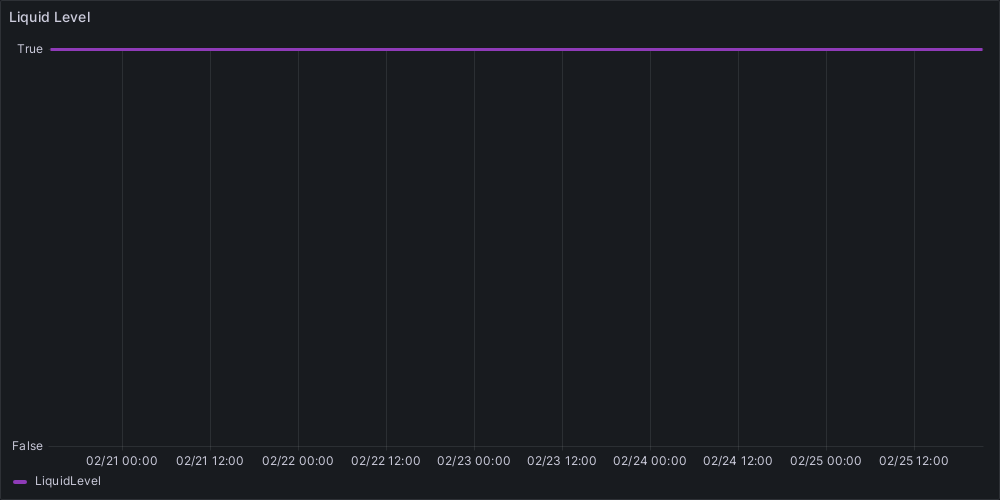}
\caption{Liquid level sensor results.}
\label{fig:liquid_level}
\vspace{0.2cm}
\centering
\includegraphics[width=1\linewidth]{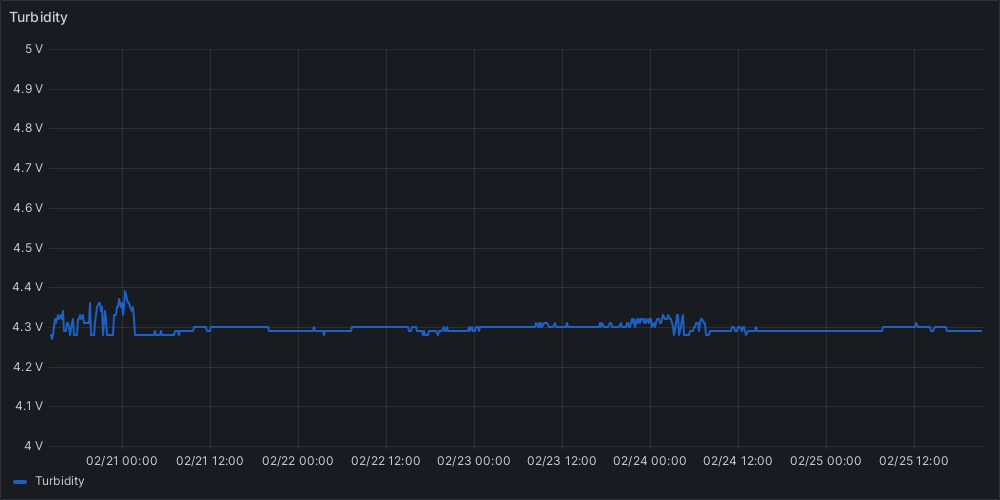}
\caption{Turbidity sensor results.}
\label{fig:turbidity}
\vspace{0.2cm}
\centering
\includegraphics[width=1\linewidth]{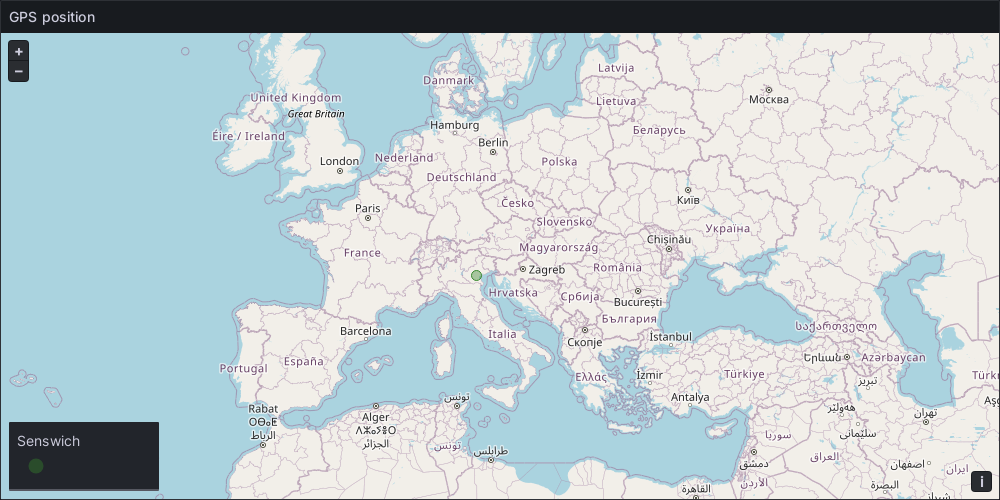}
\caption{Location of SENSWICH measured by GPS sensor.}
\label{fig:gps}
\end{figure}

Figure~\ref{fig:pH} illustrates the outcomes of the pH sensor, Figure~\ref{fig:conductivity} presents data acquired by the Electrical Conductivity Sensor, Figure~\ref{fig:temperature} corresponds to data from the Temperature sensor, Figure~\ref{fig:oxygen} to those of the Dissolved Oxygen sensor and Figure~\ref{fig:liquid_level} to those of the Liquid Level sensor, confirming the sensors' submersion in water during the specified timeframe. Additionally, Figure~\ref{fig:turbidity} relates to Turbidity.

Regarding GPS coordinates, after retrieving the payload message from AWS, we obtained accurate latitude and longitude details. Using Grafana, we effectively identified the location of SENSWICH and visually represented it on the map in Figure \ref{fig:gps}.

To measure energy consumption, we initially employed a regulator to supply 5.25~V as the input voltage to the Arduino. Table \ref{tab:power_consumption_overview} provides a comprehensive overview of power consumption across distinct sensor phases, aiding in identifying the energy utilization patterns of the Arduino MKRWAN 1310 during various phases. 
\begin{table}[h]
\centering
\caption{Energy Consumption With Using Regulator}
\label{tab:power_consumption_overview}
\begin{tabular}{|c|c|c|c|}
\hline
Phase & Duration (s) & Current (mA) & Energy (J) \\
\hline
pH & 80 & 18.2 & 12.37 \\
pH + TB & 10 & 30 & 2.55 \\
Analog Reading & 5.12 & 39 & 1.69 \\
DO & 80 & 16.5 & 11.22 \\
DO + LV & 10 & 26.1 & 2.21 \\
Analog Reading & 5.12 & 35 & 1.52 \\
EC & 20 & 24.7 & 4.19 \\
Analog Reading & 5.12 & 33.9 & 1.47 \\
LoRa TX & 2 & 41.8 & 0.71 \\
Standby & 382.64 & 15.7 & 51.06 \\
\hline
\end{tabular}
\end{table}

We can infer that Table \ref{tab:power_consumption_measurements_2} illustrates power consumption measurements based on information gathered from Table \ref{tab:power_consumption_overview}.

\begin{table}[h]
\centering
\caption{Power Consumption With Using Regulator}
\label{tab:power_consumption_measurements_2}
\begin{tabular}{|c|c|}
\hline
Parameter & Value \\
\hline
Idle power (mW) & 133.45 \\
Energy/sample (mWh) & 24.73 \\
Average current (mA) & 17.45 \\
Average power (mW) & 148.38 \\
Discharge time (h) & 733.20 \\
Discharge time (days) & 30.55 \\
\hline
\end{tabular}
\end{table}

This indicates that the device may remain active for one month with a single battery charge.

Table \ref{tab:energy_consumption_assumptions} shows the energy consumption without using regulators, assuming a 100\% efficient converter from battery voltage to 5.25~V, and Table \ref{tab:power_consumption_measurements_4} illustrates power consumption measurements based on information from Table \ref{tab:energy_consumption_assumptions}.

\begin{table}[h]
\centering
\caption{Estimation of Energy Consumption Without Using Regulator}
\label{tab:energy_consumption_assumptions}
\begin{tabular}{|c|c|c|c|}
\hline
Phase & Duration (s) & Current (mA) & Energy (J) \\
\hline
PH & 80 & 12.4 & 5.208 \\
PH+TB & 10 & 26.5 & 1.39 \\
Analog Reading & 5.12 & 38.2 & 1.02 \\
DO & 80 & 10.2 & 4.28 \\
DO+LV & 10 & 21.7 & 1.13 \\
Analog Reading & 5.12 & 33.2 & 0.89 \\
EC & 20 & 17 & 1.78 \\
Analog Reading & 5.12 & 28 & 0.75 \\
LoRa TX & 2 & 38 & 0.39 \\
Standby & 382.64 & 9.3 & 18.68 \\
\hline
\end{tabular}
\end{table}

\begin{table}[h]
\centering
\caption{Estimation of Power Consumption Without Using Regulator}
\label{tab:power_consumption_measurements_4}
\begin{tabular}{|c|c|}
\hline
Idle power (mW) & 48.82 \\
Energy/sample (mWh) & 9.87 \\
Average current (mA) & 11.28 \\
Average power (mW) & 59.267 \\
Discharge time (h) & 1552.27 \\
Discharge time (days) & 64.67 \\
\hline
\end{tabular}
\end{table}

These results suggest that the device may stay operational for up to around 64 days with a single charge, highlighting the fact that the regulator is wasting almost half of the energy, and is then not very efficient. Moreover, the Arduino standby current can be significantly reduced removing the onboard LEDs Arduino uses to signal its correct operation, further extending the battery duration.

\begin{figure}[h]
  \centering
  \subfloat[]{\includegraphics[width=0.75\linewidth]{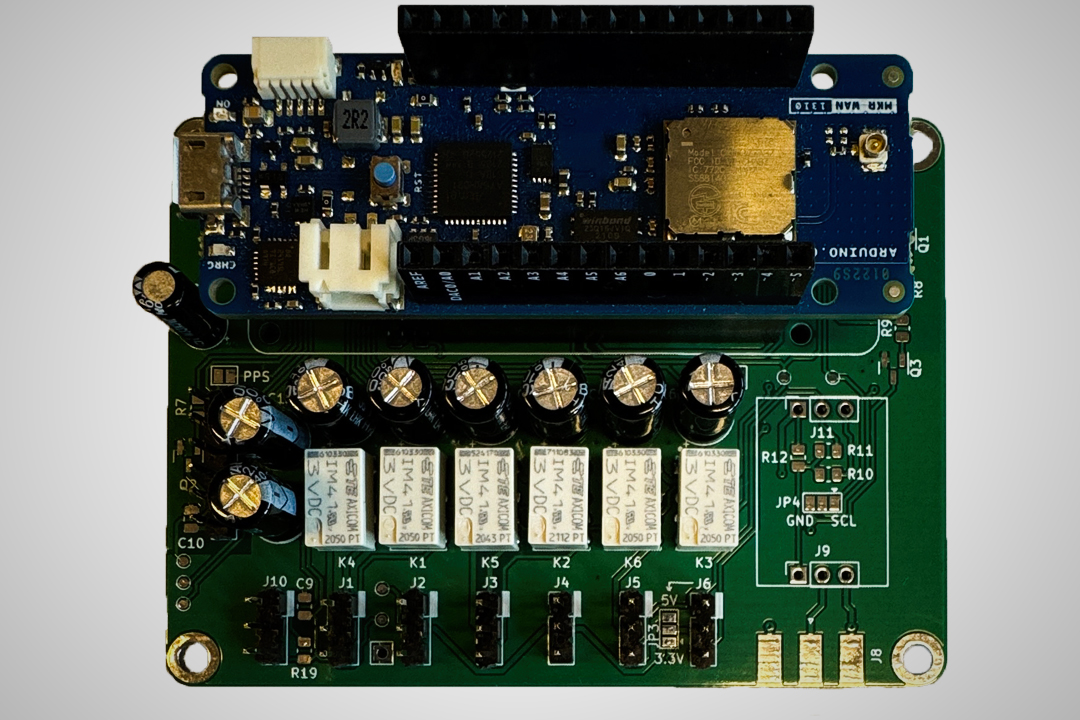}\label{fig:front_pcb}}

  \subfloat[]{\includegraphics[width=0.75\linewidth]{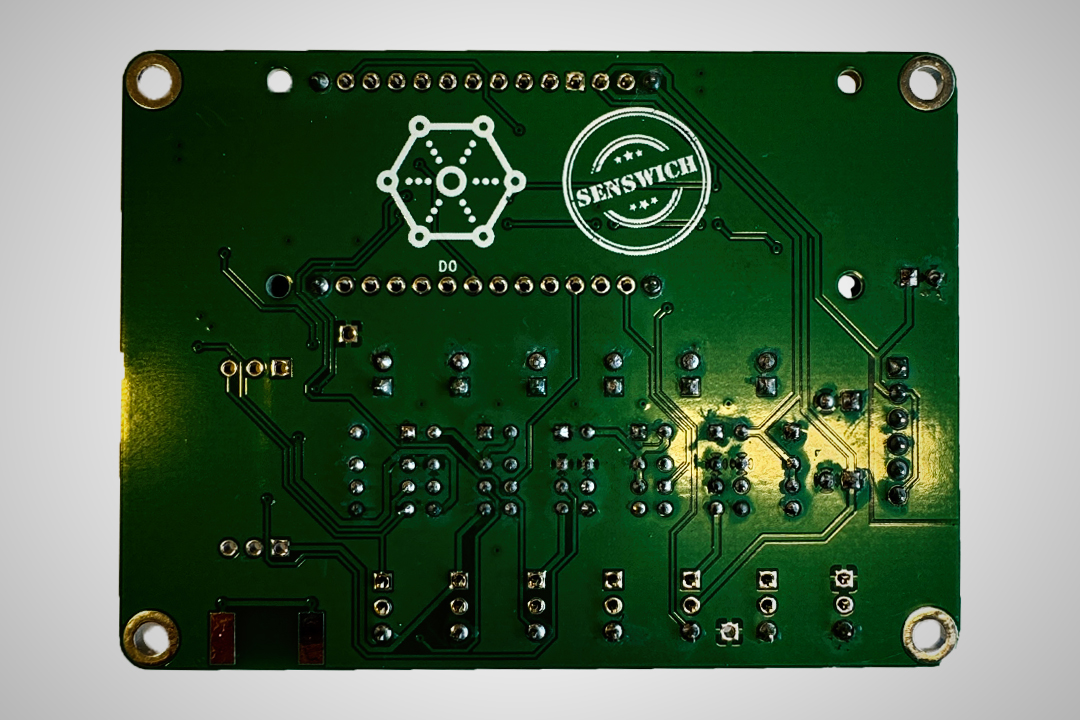}\label{fig:back_pcb}}
  \caption{Front part (a) and back part (b) of the SENSWICH PCB.}
  \label{fig:new_pcb}
\end{figure}

Given the effectiveness of the developed prototype, we have recently created the first SENSWICH PCB in order to quicly reproduce more units of the system. The first PCB unit is presented in Figure~\ref{fig:new_pcb}.

\section{Conclusion and Future Work}\label{sec:conclusion}

In analyzing the results, the combination of AWS and Grafana proves to be a powerful tool, enabling researchers to gain a comprehensive real-time view of data collected from the surface device, SENSWICH. The examination of the data reveals consistent stability in insights obtained from sensors over consecutive days, confirming the choice of the sensors. While considering a future upgrade for certain sensors, such as the electrical conductivity sensor, there is contemplation about replacing it for better outcomes. Another consideration involves the redesign of the SENSWICH floating case, while continuing to use the previous configuration ~\cite{metrosea2023} for tests in rivers and the open sea. The goal is to redesign it with an emphasis on stability on the water and resilience against harsh winds and sea storms. Noteworthy is the cost-effectiveness of this device, priced below a thousand US dollars, making it a practical alternative to some similar measuring buoys currently in use. Our attention is directed towards an in-depth investigation and the forthcoming production of additional SENSWICH units. Additionally, future plans include analyzing results in the Venetian Lagoon, with intentions to deploy it in Chioggia for further exploration.

\section*{Acknowledgments}
This project has been partially funded by the European Union under the Italian National Recovery and Resilience Plan (NRRP), as part of the NextGenerationEU initiative. The financial support is specifically attributed to the "Telecommunications of the Future" partnership (PE0000001 - program "RESTART"), the National Biodiversity Future Center (NBFC), PNRR, CN00000033, Spoke 1, and the FSE REACT EU, PON Research and Innovation 2014-2020 (DM 1062/2021). 

\bibliographystyle{IEEEtran}
\bibliography{IEEEabrv,final_ref,references}

\begin{thebibliography}{10}
\providecommand{\url}[1]{#1}
\csname url@samestyle\endcsname
\providecommand{\newblock}{\relax}
\providecommand{\bibinfo}[2]{#2}
\providecommand{\BIBentrySTDinterwordspacing}{\spaceskip=0pt\relax}
\providecommand{\BIBentryALTinterwordstretchfactor}{4}
\providecommand{\BIBentryALTinterwordspacing}{\spaceskip=\fontdimen2\font plus
\BIBentryALTinterwordstretchfactor\fontdimen3\font minus \fontdimen4\font\relax}
\providecommand{\BIBforeignlanguage}[2]{{%
\expandafter\ifx\csname l@#1\endcsname\relax
\typeout{** WARNING: IEEEtran.bst: No hyphenation pattern has been}%
\typeout{** loaded for the language `#1'. Using the pattern for}%
\typeout{** the default language instead.}%
\else
\language=\csname l@#1\endcsname
\fi
#2}}
\providecommand{\BIBdecl}{\relax}
\BIBdecl

\bibitem{natura2000}
\BIBentryALTinterwordspacing
``{European Environment Agency - The Natura 2000 protected areas network},'' last time accessed: March 2024. [Online]. Available: \url{https://www.eea.europa.eu/themes/biodiversity/natura-2000}
\BIBentrySTDinterwordspacing

\bibitem{biodiversity}
\BIBentryALTinterwordspacing
``{RESEARCHITALY - Biodiversity: national research centre established under the coordination of the CNR},'' last time accessed: March 2024. [Online]. Available: \url{https://www.researchitaly.mur.gov.it/en/biodiversity-national-research-centre-established-under-the-coordination-of-the-cnr/}
\BIBentrySTDinterwordspacing

\bibitem{eu-bio}
\BIBentryALTinterwordspacing
``Biodiversity strategy for 2030,'' last time accessed: March 2024. [Online]. Available: \url{https://environment.ec.europa.eu/strategy/biodiversity-strategy-2030_en}
\BIBentrySTDinterwordspacing

\bibitem{un-od}
\BIBentryALTinterwordspacing
``The unesco oceans decade,'' last time accessed: March 2024. [Online]. Available: \url{https://oceandecade.org/}
\BIBentrySTDinterwordspacing

\bibitem{metrosea2023}
M.~Ghalkhani, F.~Campagnaro, A.~Pozzebon, D.~D. Battisti, M.~Biagi, and M.~Zorzi, ``A {LoRaWAN} network for real-time monitoring of the venice lagoon: Initial assessments,'' in \emph{IEEE International Workshop on Metrology for the Sea; Learning to Measure Sea Health Parameters (MetroSea)}, 2023.

\bibitem{garcia86}
E.~Garcia, ``A smart low-power meteorological system,'' in \emph{Proc. IEEE/MTS OCEANS}, Washington DC, US, Sep. 1986.

\bibitem{tabs}
{N. L. Guinasso et al.}, ``{Observing and forecasting coastal currents: Texas Automated Buoy System (TABS)},'' in \emph{Proc. IEEE/MTS OCEANS}, Honolulu, HI, USA, Nov. 2001.

\bibitem{themo}
R.~Diamant, A.~Knapr, S.~Dahan, I.~Mardix, J.~Walpert, and S.~DiMarco, ``{THEMO: The Texas A\&M - University of Haifa - Eastern Mediterranean Observatory},'' in \emph{Proc. IEEE/MTS OCEANS}, Kobe, Japan, May 2018.

\bibitem{catania_sensore}
F.~Busacca, L.~Galluccio, S.~Mertens, D.~Orto, S.~Palazzo, and S.~Quattropani, ``An experimental testbed of an internet of underwater things,'' in \emph{Proc. ACM WiNTECH}, London, UK, Sep. 2020.

\bibitem{petrioliUcomms}
C.~Petrioli, R.~Petroccia, D.~Spaccini, A.~Vitaletti, T.~Arzilli, D.~Lamanna, A.~Galizial, and E.~Renzi, ``The sunrise gate: Accessing the sunrise federation of facilities to test solutions for the internet of underwater things,'' in \emph{IEEE UComms)}, 2014.

\bibitem{subcultron}
\BIBentryALTinterwordspacing
``{subCULTron - Submarine Cultures Perform Long-term Robotic Exploration of Unconventional Environmental Niches},'' last time accessed: March 2024. [Online]. Available: \url{https://labust.fer.hr/labust/research/projects/subcultron}
\BIBentrySTDinterwordspacing

\bibitem{h2orobotics}
\BIBentryALTinterwordspacing
{H2ORobotics}. H2orobotics website. [Online]. Available: \url{https://h2o-robotics.com/}
\BIBentrySTDinterwordspacing

\bibitem{datasheet}
\BIBentryALTinterwordspacing
{IM Relays}. Im relays data sheet. [Online]. Available: \url{https://datasheetspdf.com/pdf-file/1259629/TE/IM41TS/1}
\BIBentrySTDinterwordspacing

\bibitem{aws-cloud}
\BIBentryALTinterwordspacing
``{AWS Cloud},'' {Last time accessed: March 2024}. [Online]. Available: \url{https://aws.amazon.com/}
\BIBentrySTDinterwordspacing

\bibitem{iot-core-lorawan}
\BIBentryALTinterwordspacing
``{AWS IoT Core for LoRaWAN},'' {Last time accessed: March 2024}. [Online]. Available: \url{https://aws.amazon.com/iot-core/lorawan/}
\BIBentrySTDinterwordspacing

\bibitem{cayenne-lpp}
\BIBentryALTinterwordspacing
``{CayenneLPP},'' {Last time accessed: March 2024}. [Online]. Available: \url{https://www.thethingsnetwork.org/docs/devices/arduino/api/cayennelpp/}
\BIBentrySTDinterwordspacing

\bibitem{mqtt-protocol}
\BIBentryALTinterwordspacing
``{MQTT Protocol},'' {Last time accessed: March 2024}. [Online]. Available: \url{https://mqtt.org/}
\BIBentrySTDinterwordspacing

\bibitem{iot-core}
\BIBentryALTinterwordspacing
``{AWS IoT Core},'' {Last time accessed: March 2024}. [Online]. Available: \url{https://aws.amazon.com/iot-core/}
\BIBentrySTDinterwordspacing

\bibitem{cups}
\BIBentryALTinterwordspacing
``{Configuration and Update Server ({CUPS})},'' {Last time accessed: March 2024}. [Online]. Available: \url{https://www.thethingsindustries.com/docs/gateways/concepts/lora-basics-station/cups/}
\BIBentrySTDinterwordspacing

\bibitem{lns}
\BIBentryALTinterwordspacing
``{{LoRaWAN} Network Server (LNS)},'' {Last time accessed: March 2024}. [Online]. Available: \url{https://www.thethingsindustries.com/docs/gateways/concepts/lora-basics-station/lns/}
\BIBentrySTDinterwordspacing

\bibitem{lorawan-architecture}
\BIBentryALTinterwordspacing
``{LoRaWAN network architecture},'' {Last time accessed: March 2024}. [Online]. Available: \url{https://docs.aws.amazon.com/whitepapers/latest/implementing-lpwan-solutions-with-aws/lorawan-network-architecture.html}
\BIBentrySTDinterwordspacing

\bibitem{basics-station}
\BIBentryALTinterwordspacing
``{How to Use {LoRa} Basics Station},'' {Last time accessed: March 2024}. [Online]. Available: \url{https://lora-developers.semtech.com/documentation/tech-papers-and-guides/how-to-use-lora-basics-station}
\BIBentrySTDinterwordspacing

\bibitem{iot-rules}
\BIBentryALTinterwordspacing
``{Rules for {AWS IoT}},'' {Last time accessed: March 2024}. [Online]. Available: \url{https://docs.aws.amazon.com/iot/latest/developerguide/iot-rules.html}
\BIBentrySTDinterwordspacing

\bibitem{aws-lambda}
\BIBentryALTinterwordspacing
``{What is {AWS} Lambda?}'' {Last time accessed: March 2024}. [Online]. Available: \url{https://docs.aws.amazon.com/lambda/latest/dg/welcome.html}
\BIBentrySTDinterwordspacing

\bibitem{aws-timestream}
\BIBentryALTinterwordspacing
``{What is Amazon Timestream?}'' {Last time accessed: March 2024}. [Online]. Available: \url{https://docs.aws.amazon.com/timestream/latest/developerguide/what-is-timestream.html}
\BIBentrySTDinterwordspacing

\bibitem{aws-ec2}
\BIBentryALTinterwordspacing
``{Amazon Elastic Compute Cloud Documentation},'' {Last time accessed: March 2024}. [Online]. Available: \url{https://docs.aws.amazon.com/ec2/}
\BIBentrySTDinterwordspacing

\bibitem{grafana}
\BIBentryALTinterwordspacing
``{Grafana: The open observability platform | Grafana Labs},'' {Last time accessed: March 2024}. [Online]. Available: \url{https://grafana.com/}
\BIBentrySTDinterwordspacing

\end{thebibliography}

\end{document}